\documentclass[showpacs, secnumarabic, amssymb, nobibnotes, aps, prb, reprint]{revtex4-1}
\usepackage{graphicx}%
\usepackage{dcolumn}%
\usepackage{bm}%
\usepackage{lineno}
\usepackage{epstopdf,color}

\begin{document}
\title{Anomalous Hall Effect in \\ Ge$_{1\textrm{-}x\textrm{-}y}$Pb$_{x}$Mn$_{y}$Te Composite System}

\author{A.~Podg\'{o}rni}
\email[Electronic mail: ]{podgorni@ifpan.edu.pl}
\author{L.~Kilanski}
\author{W.~Dobrowolski}
\author{M. G\'{o}rska}
\author{\framebox{V. Domukhovski}}
\author{B.~Brodowska}
\author{A.~Reszka}
\author{B.J. Kowalski}
\affiliation{Institute of Physics, Polish Academy of Sciences, al. Lotnikow 32/46, 02-668 Warsaw, Poland}

\author{V.~E.~Slynko}
\author{E.~I.~Slynko}
\affiliation{Institute of Materials Science Problems, Ukrainian Academy of Sciences, 5 Wilde Street, 274001 Chernovtsy, Ukraine}

\date{\today}

\begin{abstract}
The purpose of this study was to investigate the magnetotransport properties of the Ge$_{0.743}$Pb$_{0.183}$Mn$_{0.074}$Te mixed crystal. The results of magnetization measurements indicated that the compound is a spin-glass-like diluted magnetic semiconductor with critical temperature $T_{SG}=97.5$~K. Nanoclusters in the sample are observed. Both, matrix and clusters are magnetically active. Resistivity as a function of temperature has a minimum at 30 K. Below the minimum a variable-range hopping is observed, while above the minimum a metallic-like behavior occurs. The crystal has high hole concentration, $p = 6.6 \times 10^{20}$ cm$^{-3}$, temperature-independent. Magnetoresistance amplitude changes from -0.78 to 1.18 \% with increase of temperature. In the magnetotransport measurements we observed the anomalous Hall effect (AHE) with hysteresis loops. Calculated AHE coefficient, $R_{S} = 2.0 \times 10^{6}~$m$^{3}$/C, is temperature independent. The analysis indicates the extrinsic skew scattering mechanism to be the main physical mechanism responsible for AHE in Ge$_{0.743}$Pb$_{0.183}$Mn$_{0.074}$Te alloy.

\end{abstract}





\maketitle


Transition metal doped IV-VI compounds are a subject in the recent years of a significant interest  due to the presence of the carrier mediated ferromagnetism\cite{RKKY} with Curie temperature, $T_{C}$, about 190 K in Ge$_{0.92}$Mn$_{0.08}$Te thin films.\cite{Fukuma08a} The nanocomposite crystals with clusters of magnetic impurities might lead to a further increase of $T_{C}$, above 190$\;$K. GePbMnTe mixed crystals create opportunity to control electric and magnetic properties independently.\cite{Asada08a} Moreover, in GeTe-based semimagnetic semiconductors the anomalous Hall effect has been observed.\cite{Kilanski13a} This effect makes it possible to use the GeTe-based semimagnetic semiconductors for devices with electrically readable magnetic storage. Bulk mixed GePbMnTe crystals are easier and less expensive to obtain than thin layers, but their properties are not well known at present.

For the purpose of this study a single GePbMnTe ingot was grown by the modified Bridgman method.\cite{Aust58a} The energy dispersive x-ray fluorescence technique (EDXRF) was used to determine the sample chemical composition. For the present investigation the Ge$_{0.743}$Pb$_{0.183}$Mn$_{0.074}$Te sample has been chosen.
\begin{figure}[h]
  \begin{center}
    \includegraphics[width = 0.5\textwidth, bb = 0 30 850 580]
    {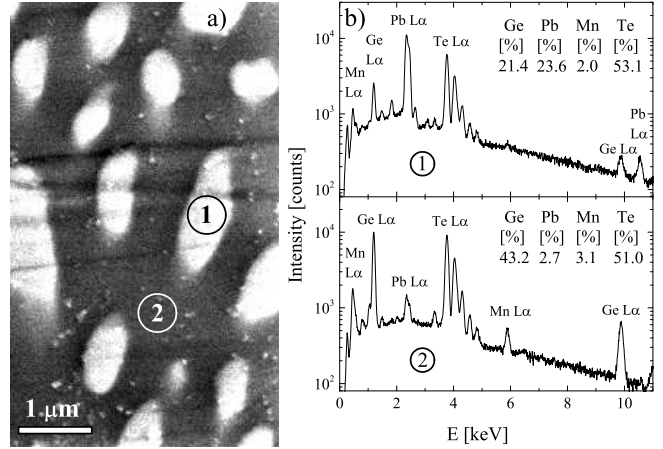}
  \end{center}
  \caption{\small The SEM image (a) and EDX spectra (b) measured at selected spots marked with circles for Ge$_{0.743}$Pb$_{0.183}$Mn$_{0.074}$Te crystal.}
  \label{fig.picture}
\end{figure}

Powder x-ray diffraction (XRD) measurements were performed at room temperature to investigate the structural properties of the sample. The obtained diffraction pattern was analyzed using the Rietveld refinement method. Studies revealed that the considered crystal is a double-phase system. The main phase is a distorted along (111) direction NaCl structure with lattice constant $a_{1} = 5.93$~\AA~and the angle of distortion $\alpha = 89.44^{\circ}$. The secondary phase (rich in  lead) is a cubic NaCl structure with lattice constant $a_{2}$ equal $6.39$~\AA. Obtained parameters are between the values for pure GeTe and PbTe crystals \cite{Galazka99a}, and for both phases the Vegard's law is fulfilled.

\begin{figure}[h]
  \begin{center}
    \includegraphics[width = 0.5\textwidth, bb = 0 50 530 580]
    {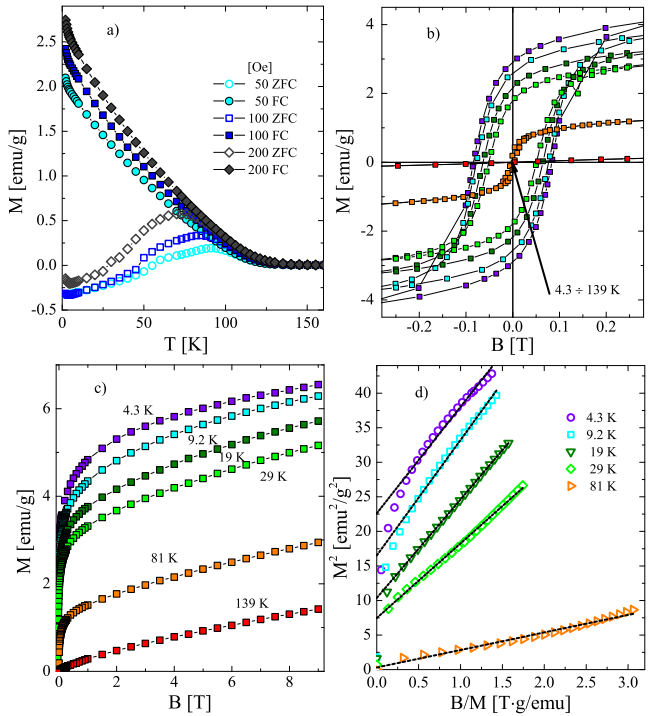}
  \end{center}
  \caption{\small Results of magnetization measurements for Ge$_{0.743}$Pb$_{0.183}$Mn$_{0.074}$Te crystal.}
  \label{fig.magnetism}
\end{figure}

To study the chemical heterogeneity of the crystal, scanning electron microscopy (SEM) combined with energy dispersive x-ray spectrometer detector (EDX) was used (see Fig. \ref{fig.picture}). Surface image shows the presence of nanoscale clusters rich in Pb. The magnetic properties of the alloy may be significantly influenced by the presence of two phases with different chemical compositions.

Figure \ref{fig.magnetism} presents the magnetic properties of the Ge$_{0.743}$Pb$_{0.183}$Mn$_{0.074}$Te crystal measured by Quantum Design XL-5 magnetometer and Weiss extraction method implemented in  LakeShore 7229 AC Susceptometer/DC Magnetometer system. Magnetization curves $M(T)$ (Fig. \ref{fig.magnetism}a) indicate a presence of magnetic order below 100 K. The difference between the field cooled (FC) and zero-field cooled (ZFC) M(T) curves indicates that we do not observe a typical paramagnet-ferromagnet transition. The nearly square-shaped hysteresis loops with rather high coercive fields (Fig. \ref{fig.magnetism}b) are a sign that the domain structure is present in the material. However, the lack of saturation of the $M(B)$ curves at $B>2$~T (Fig. \ref{fig.magnetism}c) is a signature of a large frustration in our system and a formation of a mixed ferromagnet-spin-glass-like state.\cite{Kilanski10a,Podgorni12a} The Arrot plots presented in Figure \ref{fig.magnetism}d show that besides frustration ferromagnetic interactions occur below the magnetic transition temperature. The approximation of the maximum position in the $M(T)|_{B\rightarrow0}$ curves was used to estimate the spin-glass-like transition temperature, $T_{SG}$, as about 97.5 K. This transition we assign to the Ge-rich GePbMnTe phase. Below 5 K in the FC $M(T)$ curves a signature of a magnetization coming from the Pb-rich phase (possibly paramagnetic) is observed.

\begin{figure*}[ht]
  \begin{center}
    \includegraphics[width = \textwidth, bb = 0 230 820 580]
    {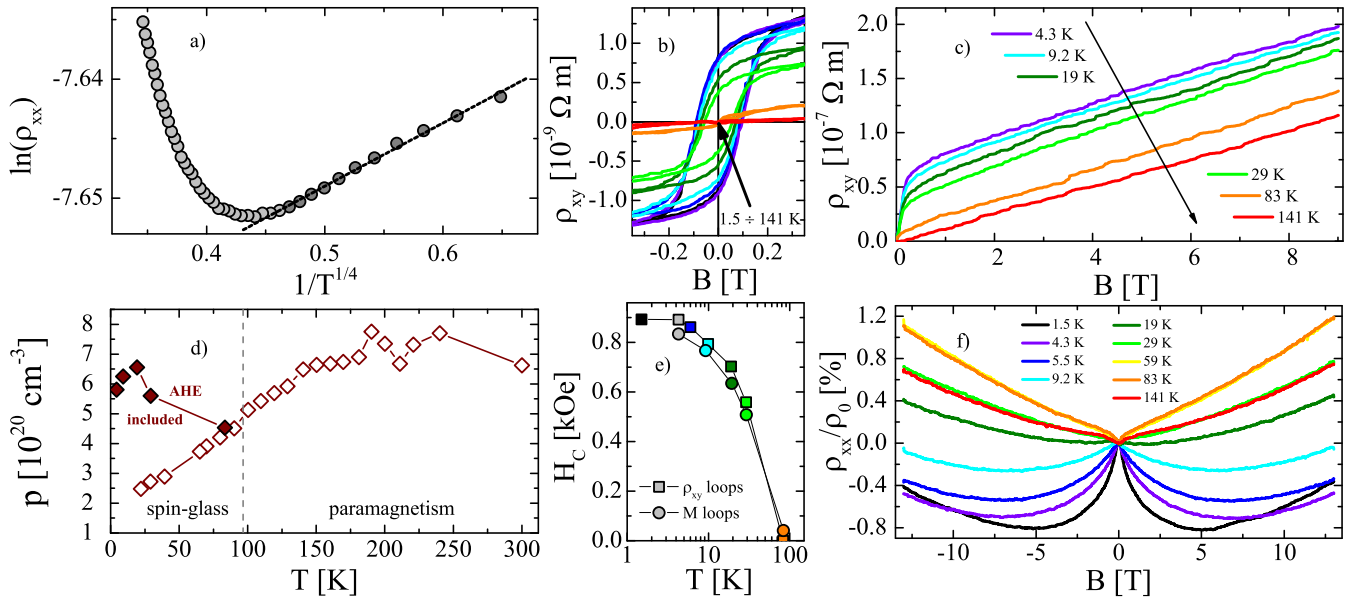}\\
  \end{center}
  \caption{\small Results of magnetotransport measurements: (a) temperature dependence of resistivity with $\rho_{xx}=\rho_{0}~exp[(T_{0}/T)^{1/4}]$ function fitted, (b) $\rho_{xy}$ hysteresis loops at selected temperatures, (c) Hall resistivity as a function of magnetic field at selected temperatures (d) carrier concentration calculated as $p$$\,$$=$$\,$$(R_{H}e)^{-1}$ and by using Eq.\ref{math.ahe} (open and full symbols respectively), (e) temperature dependence of the coercivity field estimated from resistivity (squares) and magnetization (circles) hysteresis, (f) magnetoresistance as a function of magnetic field at selected temperatures.}
  \label{fig.transport}
\end{figure*}

Standard six-contact dc-current Hall geometry with magnetic field $B<13$~T  was used to determine the magnetotransport properties of our Ge$_{0.743}$Pb$_{0.183}$Mn$_{0.074}$Te crystal. The temperature dependence of the resistivity, $\rho_{xx}(T)$, was measured. The sample exhibits metallic-like behavior typical for degenerate semiconductors, but below 30 K a minimum in resistivity is observed. We note that the change of $\rho_{xx}$ over the whole range of temperatures is relatively small, and the ratio of values obtained at the minimum and room temperature is close to 26 \%. The low-temperature dependence of $\rho_{xx}(T)$ was fitted with the Mott's law
\begin{equation}
\rho_{xx}=\rho_{0} \cdot exp [(T_{0}/T)^{1/4}],
\label{math.mott}
\end{equation}
where $\rho_{0}$ is the residual resistance and $T_{0}$ is a constant inversely proportional to the density of states at the Fermi level.\cite{Shklovskii84a} The result of fitting (line) is shown in Figure \ref{fig.transport}a. We obtained a good agreement of our data with the Mott's law, typical for variable-range hopping.\cite{Shklovskii84a, Chen07a}

The $\rho_{xy}(B)$ dependence was measured from 300 K down to 4.5 K. The results were used to calculate of the temperature dependence of the Hall carrier concentration $p$$\,$$=$$\,$$(R_{H}e)^{-1}$ (where $R_{H}=\rho_{xy}/B$) in the paramagnetic temperature region. As we can see in Figure \ref{fig.transport}d the Hall carrier concentration (open symbols) above $T_{SG}$ is temperature independent. Such situation is typical for degenerate semiconductors. In the spin-glass region the anomalous Hall effect has a significant impact on the measured $R_{H}$ value and as a consequence we see an apparent decrease in the concentration (see also Ref. \onlinecite{Kilanski09a}). The influence of the AHE has to be taken into account for a proper estimation of the carrier concentration in the spin-glass-like state.

The magnetic field dependence of the Hall resistivity $\rho_{xy}(B)$ (see Fig. \ref{fig.transport}c) indicates a presence of a strong anomalous Hall effect at every measured temperature below the magnetic transition temperature, $T_{SG}$. In case of materials showing AHE behavior, the Hall resistivity can be described in the following form:
\begin{equation}
\rho_{xy} = R_{H} \cdot B + \mu_{0} \cdot R_{S} \cdot M,
\label{math.ahe}
\end{equation}
where $R_{H}$ is the normal Hall constant, $\mu_{0}$ is vacuum permeability, and $R_{S}$ is the anomalous Hall constant. The first term in Eq.\ref{math.ahe} describes the normal Hall effect caused by Lorentz force, and the second the contribution of AHE due to the asymmetric carrier scattering. The magnetotransport data together with the previously measured magnetization at the same selected temperatures let us to calculate both of the Hall constants. We obtained $R_{H} = 1.1 \times 10^{8}$~m$^{3}$/C and $R_{S} = 2.0 \times 10^{6}~$m$^{3}$/C at liquid helium temperature. Both constants are nearly temperature independent. Obtained parameters are similar to those reported in our former study of GeMnTe-SnMnTe\cite{Kilanski13a}.

Taking into account the $R_{H}$ values estimated from the AHE analysis we obtain a corrected carrier concentration at temperatures below magnetic transition. The calculated concentration, $p$, for spin-glass region is presented in Figure \ref{fig.transport}d by full symbols. We found that the Ge$_{1-x-y}$Pb$_{x}$Mn$_{y}$Te crystal is a $p$-type semiconductor with temperature independent (over the investigated temperature range) high carrier concentration $p = 6.6 \times 10^{20}$~cm$^{-3}$ (for 300 K).

In the literature two major semiclassical mechanisms explaining the AHE are known - skew scattering and side jump. For the skew scattering\cite{Smit55a} the $\rho_{xy}\propto\rho_{xx}$, while for the side jump\cite{Berger70a} the $\rho_{xy}\propto\rho_{xx}^{2}$. A detailed analysis of AHE includes a scaling relationship described by the following equation:
\begin{equation}
\rho_{xy}(B) = R_{H} \cdot B + c_{H} \cdot \rho_{xx}^{n_{H}} \cdot M,
\label{math.ahe2}
\end{equation}
where $c_{H}$ and $n_{H}$ are scaling parameters, which give us information about a dominant scattering mechanism leading to AHE in the sample. For liquid helium temperature we obtained $n_{H} = 1.09$. The $n_{H}$ is nearly temperature independent. The calculated value indicates that the extrinsic skew scattering is a dominant mechanism responsible for AHE in studied Ge$_{0.743}$Pb$_{0.183}$Mn$_{0.074}$Te crystal. The main role of this mechanism was also pointed out for bulk Ge$_{1-x-y}$Sn$_{x}$Mn$_{y}$Te crystals.\cite{Kilanski13a}

Low-field $\rho_{xx}(B)$ curves show hysteresis loops of AHE. As presented in Figure \ref{fig.transport}b, the hysteresis loops are observed up to the critical temperature. The presence of hysteresis indicates a large frustration in our spin-glass-like system. In Figure \ref{fig.transport}e we gathered the values of the coercivity field, $H_{C}$, estimated from AHE loops (squares) and from magnetization curves (circles) as a function of temperature. As we can see both quantities are in good agreement and fall down to zero at the transition temperature, $T_{SG}$. This indicates a presence of asymmetric carrier scattering in our spin-glass-like system, directly correlated with magnetic properties of the alloy.

The magnetoresistance was measured together with the Hall effect. At low-temperatures a negative magnetoresistance with a minimum at about 0.5 T was observed. Contribution of the negative magnetoresistance to the total magnetoresistance decreases with increase of temperature. At high temperatures (T $>$ $T_{SG}$) only the positive magnetoresistance proportional to $B^{2}$ was observed. The $\rho_{xx}(B)$ changes from  -0.78 up to 1.18 \% of  $\rho_{xx}(B = 0)$ with growing temperature, till $T_{SG}$ is reached. Above the magnetic transition temperature the trend is opposite, the values of magnetoresistance become lower. The observed negative magnetoresistance may be due to the weak localization.

To conclude, we have shown results of magnetotransport measurements for a bulk spin-glass-like Ge$_{0.743}$Pb$_{0.183}$Mn$_{0.074}$Te crystal.

Temperature-dependent resistivity indicates a metalic-like trend, but a minimum below 30 K is observed. The analysis of $\rho_{xx}(T)$ pointed to the variable-range hopping. The crystal is a $p$-type semiconductor with high Hall carrier concentration $p = 6.6 \times 10^{20}$~cm$^{-3}$.

The anomalous Hall effect with wide hysteresis loops was observed. Calculated AHE coefficient, $R_{S}$, is temperature independent, with the value close to $2.0 \times 10^{6}$~m$^{3}$/C. A detailed analysis of the effect indicates that the main physical mechanism responsible for AHE in Ge$_{0.743}$Pb$_{0.183}$Mn$_{0.074}$Te alloy is the extrinsic skew scattering mechanism.

Magnetotransport properties of the bulk Ge$_{0.743}$Pb$_{0.183}$Mn$_{0.074}$Te crystal are strongly correlated with the spin-glass-like nature of the alloy.

\section*{Acknowledgments}

\noindent  The research was supported by the Foundation for Polish Science - POMOST/2011-4/2 Programme co-financed by the European Union within European Regional Development Fund.


\begin{thebibliography}{natbib}

\bibitem{RKKY} M.A.~Ruderman and C.~Kittel, \emph{Phys. Rev.} \textbf{96}, 99 (1954). \\ T.~Kasuya, \emph{Prog. Theor. Phys.} \textbf{16}, 45 (1956). \\ K.~Yosida, \emph{Phys. Rev.} \textbf{106}, 893 (1957).
\bibitem{Fukuma08a} Y. Fukuma et al., {\em Appl. Phys. Lett.} {\bf 93}, 252502 (2008).
\bibitem{Asada08a} H. Asada et al., {\em IEEE} {\bf 44}, 2696 (2008).
\bibitem{Kilanski13a} L. Kilanski, et al., {\em J. Appl. Phys.} {\bf 113}, 063702 (2013).
\bibitem{Aust58a} K. T. Aust and B. Chalmers, {\em Can. J. phys.} {\bf 36}, 977 (1958).
\bibitem{Galazka99a} R. R. Galazka et al., {\em Landolt-B$\ddot{o}$rnstein, New Series, Group III/41, chapter Semiconductors}, Berlin, Heidelberg, Spriner-Verlag, 1999.
\bibitem{Kilanski10a} L. Kilanski, et al., {\em Phys. Rev. B} {\bf 82}, 094427 (2010).
\bibitem{Podgorni12a} A. Podg\'{o}rni, et al., {\em Acta. Phys. Pol. A} {\bf 122}, 1012 (2012).
\bibitem{Shklovskii84a} B. I. Shklovskii and A. L. Efros, {\em Electronic Properties of Doped Semiconductors}, Berlin, Spriner, 1984.
\bibitem{Chen07a} W. Q. Chen et al., {\em Appl. Phys. Lett.} {\bf 90}, 142514 (2007)
\bibitem{Kilanski09a} L. Kilanski, et al., {\em J. Appl. Phys.} {\bf 105}, 103901 (2009)
\bibitem{Smit55a} J. Smit et al., {\em Physica (Amsterdam)} {\bf 12}, 877 (1955)
\bibitem{Berger70a} L. Berger et al., {\em Phys. Rev. B} {\bf 2}, 4559 (1970)


\end{thebibliography}
\end{document}